\documentclass[prb,twocolumn,superscriptaddress,showpacs,amsmath,amssymb]{revtex4-1}

\usepackage{graphicx}
\usepackage{latexsym}
\usepackage{amsmath}
\usepackage{amssymb}
\usepackage{amsfonts}
\usepackage{color}
\usepackage{bm}% bold math
\usepackage{verbatim}
\usepackage{pagecolor,lipsum}

\begin{document}

 \title{ $\theta_0$ thermal  Josephson junction. }

\date{\today}

 \author{M.A.~Silaev}
 \affiliation{Department of
Physics and Nanoscience Center, University of Jyv\"askyl\"a, P.O.
Box 35 (YFL), FI-40014 University of Jyv\"askyl\"a, Finland}

%%% abstract
 \begin{abstract}
  We predict the thermal counterpart of the anomalous Josephson effect in 
  superconductor/ferromagnet/superconductor junctions
 with non-coplanar magnetic texture. 
 The heat current through the junction is shown to have the 
 phase-sensitive interference component proportional to $\cos(\theta - \theta_0)$,  
 where $\theta$ is the Josephson phase difference  and $\theta_0$ is the
 texture-dependent phase shift.
  In the generic tri-layer magnetic structure with the spin-filtering tunnel barrier
 $\theta_0$ is determined by the spin chirality of magnetic configuration and can be considered as the direct manifestation
  of the energy transport with participation of spin-triplet Cooper pairs.  
 In case of the ideal spin filter  the phase shift is shown to be  robust against spin relaxation caused by the spin-orbital 
  scattering.  Possible applications of the coupling between heat flow and magnetic precession are discussed.  
   \end{abstract}

%%% PACS numbers
 \maketitle

During the recent years large attention has been devoted to the 
emerging field of phase-coherent calortironics in hybrid superconducting structures\citep{Martinez-Perez2014}.
The mechanism of phase-sensitive heat transport is based on the thermal counterpart of the Josephson effect \cite{Maki1965,Guttman1997,Guttman1998,Zhao2003,Zhao2004}
which occurs in the system consisting of two superconductors S$_1$ and  S$_2$
separated a weak link and residing at temperatures 
$T_1$ and $T_2$, respectively. The non-zero temperature bias (for definiteness we assume
that  $T_1>T_2$) generates a stationary heat flow from 
S$_1$ to  S$_2$ given by the heat current-phase relation (HCPR)
 \begin{equation}\label{Eq:HCPR}
 \dot Q_{tot}(T_1,T_2,\theta)  = \dot Q_{qp} - \dot Q_{int}\cos\theta \,,
 \end{equation}  
where $\theta$ is the phase difference between  superconducting  electrodes.
Here the first term is the usual quasiparticle heat current while the second one 
describes the contribution of energy transfer with participation of Cooper pairs. 
In accordance with Onsager symmetry the heat current is time-reversal invariant since 
the phase-coherent term in Eq.(\ref{Eq:HCPR}) does not change under the phase inversion
$\dot Q_{tot}(\theta) = \dot Q_{tot}(-\theta)$.

%%%%%%%%%%%%%%%%%%%%%%%%%%%%%%%%%%%%%%%%
Experimentally the interplay of heat transport and Josephson phase difference has been studied starting from the observations
of thermoelectric effects in superconductor/ normal metal/superconductor junctions \cite{Ryazanov1981,Ryazanov1982,Kartsovnik,Panaitov1984,Kaplunenko1985,Logvenov1986}.  
Recently the existence of coherent thermal currents (\ref{Eq:HCPR}) has been confirmed in experiments
using the Josephson heat interferometry with tunnel contacts\cite{Giazotto2012,JoseMartinez-Perez2014,Martinez-Perez2014}. 
Subsequently the number of possible applications has been suggested including
heat interferometers \cite{JoseMartinez-Perez2014, Giazotto2012, Fornieri2016a, Guarcello2016} , diodes \cite{Martinez-Perez2015}, 
transistors \cite{Fornieri2016a,DAmbrosioa2015,Giazotto2014a}, phase-tunable ferromagnetic Josephson valves\cite{GiazottoBergeret2013,Bergeret2013} and the probes of topological Andreev bound states\cite{Sothmann2016}. 
The direction $ \dot Q_{int}$ in Eq. (\ref{Eq:HCPR}) can be controlled in experiments 
providing the realization of $0-\pi$ thermal Josephson junction \cite{Fornieri2017}.   

 In the present paper we report on the possibility to obtain the generalized HCPR of the form
 \begin{equation} \label{Eq:HCPRgen}
 \dot Q_{tot}(T_1,T_2,\theta)  = \dot Q_{qp} - \dot Q_{int} \cos(\theta - \theta_0) \,,
  \end{equation} 
 which  can have an arbitrary phase-shift $\theta_0$ in contrast to the Eq.(\ref{Eq:HCPR}) studied in all previous works 
 \cite{Maki1965,Guttman1997,Guttman1998,Zhao2003,Zhao2004}. 
 This effect takes place in the systems with broken time-reversal and chiral symmetries such as the 
 S/F/S junctions with non-coplanar  magnetic textures or spin-orbital interaction. 
 It can be considered as the thermal counterpart of the anomalous Josephson effect 
 characterized by the generalized current-phase relation 
 (CPR) 
 \cite{Braude2007, Buzdin2008, Zazunov2009, Brunetti2013,Konschelle2009, Grein2009, Liu2010, Yokoyama2014,
 Margaris2010,Alidoust2013, Mironov2015, Kulagina2014,
 Moor2015, Moor2015a, Mironov2015a, Konschelle2009, Bergeret2015, 2016arXiv160707794S, Nesterov2016,Bobkova2016,Szombati2016, Silaev2017}     
 \begin{equation} \label{Eq:CPRgen}
 I (\varphi)= I_{c}\sin(\theta-\varphi_0)\; .  
 \end{equation}
 Here $I_c$ is the critical current and $\varphi_0$ is an arbitrary phase shift which however in the general 
  case is different from that in the generalized HCPR $\theta_0\neq\varphi_0$.

  %%%%%%%%%%%%%%%%%%%%%%%%%%%%%%%%%%%%%%%%%%%%%%%%%%%%
 \begin{figure}[h!]
 \centerline{$
 \begin{array}{c}
 \includegraphics[width=0.9\linewidth]{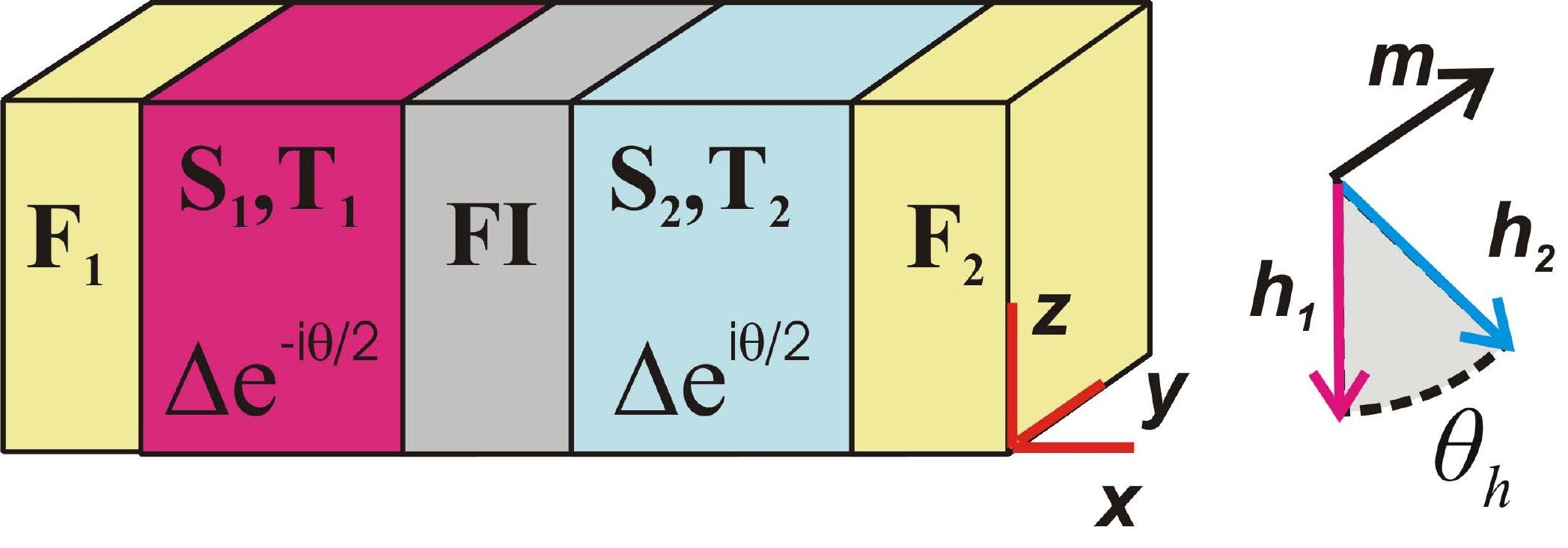} 
 \end{array}$}
 \caption{\label{Fig:SFSmodel} (Color online) 
 The sketch of FS-FI-SF system under the thermal bias with the superconducting electrodes S$_{1,2}$
 residing at different temperatures $T_{1,2}$. The exchange fields ${\bm h_1}$ and ${\bm h_2}$ in ferromagnetic electrodes 
 $F_1$ and $F_2$ form a non-coplanar system with the spin polarization ${\bm m}$ of the ferromagnetic 
 barrier (FI) .  }
 \end{figure}
 %%%%%%%%%%%%%%%%%%%%%%%%%%%%%%%%%%%%%%%%%%%%%%%%%%%%     
  %
  We demonstrate the phase-shifted HCPR (\ref{Eq:HCPRgen}) using a generic example of Josephson
 spin-valve\cite{Bergeret2013,Bergeret2014,GiazottoBergeret2013,Giazotto2015}  that contains three non-coplanar  magnetic   
 vectors, see Fig.(\ref{Fig:SFSmodel}. It consists of two  ferromagnetic layers (F) with exchange fields ${\bm h}_{1,2}$
  interacting with the superconducting electrodes (S),
separated by the spin-filter barrier with the magnetic polarization directed along ${\bm m}$. 
Recently the spin-filter effect in superconductor/ferromagnet structures 
has been demonstrated by using ferromagnetic insulators  (FI) europium chalcogenides\cite{Hao1990,Santos2008,Miao2009,Li2013,Wolf2014,2017arXiv170407241} or 
GdN tunnelling barriers\cite{Senapati2011}. 
  The role of outer  F$_{1,2}$ contacts is to induce effective exchange fields in the superconducting electrodes. 
 In case of metallic ferromagnets this can be achieved
 through the inverse the proximity effect\citep{Tokuyasu1988,Bergeret2004,Bergeret2005,Moodera1988}. 
 Alternatively F$_{1,2}$ can be ferromagnetic insulators
 and induce the effective exchange field in S$_{1,2}$ as a result of the spin-mixing scattering of conduction electrons
 \cite{Eschrig2015}.  
  %
  
 %{\bf Tunnelling heat current through the spin-filter barrier.} 
 To calculate the currents across spin-filtering barriers are we use generalized  Kuprianov-Lukichev boundary conditions \cite{KL}, 
 that include  spin-polarized 
 tunnelling at the SF interfaces \cite{Bergeret2012,Eschrig2015}. The matrix tunnelling current
 from S$_1$ to S$_2$ is given by  
 \begin{equation}
 \check I_{12}=[\check\Gamma  \check g_{1} \check\Gamma^\dagger, \check g_{2} ],
 \label{Eq:KupLuk}
 \end{equation} 
 where $\check g_{k}$ for $k=1,2$ are the matrix Green's functions (GF) in the superconducting electrodes S$_k$. 
 The spin-polarized tunnelling matrix has the form $\check\Gamma= t_+\hat\sigma_0\hat\tau_0 + t_-(\bm { m \hat\sigma})\hat\tau_3 $, 
 where ${\bm m}$ is the direction of  barrier  magnetization, $t_\pm = \sqrt{( 1\pm\sqrt{1-P^2} )/2} $ and $P$ 
 being the spin-filter efficiency of the barrier that ranges from $0$ (no polarization)  to $1$ (100\% filtering efficiency).
  The matrix GF is given by 
 $\check{g} = \left(%
 \begin{array}{cc}
  g^R & g^K \\
  0 &  g^A \\
 \end{array}%
 \right)$, 
 where $g^K$ is the  Keldysh component
 and $g^{R(A)}$ is the retarded (advanced) GF determined by the equation \cite{Bergeret2005}
 \begin{equation} \label{Eq:Usadel}
   [i\varepsilon \tau_3-i({\bm{ h\cdot S})}\tau_3 - \check{\Delta} + \check\Sigma_{s}, \check{g}] =0
 \end{equation}  
  Here $\varepsilon$ is the energy, $ \check{\Delta}=\Delta\tau_1 e^{i\tau_3\varphi}$ 
   is the order parameter with the amplitude $\Delta$ and phase $\varphi$,
   ${\bm h } $ is the exchange field,
    ${\bm S}= (\sigma_1,\sigma_2,\sigma_3)$,  $\sigma_{1,2,3}$ and $\tau_{1,2,3}$ are the Pauli matrices in spin and Nambu spaces respectively. 
   We include the spin-orbital (SO) scattering process which lead to the spin relaxation described by  \cite{Abrikosov1961,Bergeret2005}
   $\check\Sigma_{s} = ({\bm {S}}\cdot\check{g} {\bm {S}}) /8\tau_{so} $, where $\tau_{so}$ is the SO scattering time. 
 Due to the  normalization condition $\check g^2=1$ the Keldysh component can be written as
 $ g^K= (g^R - g^A)f_L$, where 
  $f_L=f_L(\varepsilon)$ is the distribution function. We assume that it has an
  equilibrium form $f^{(1,2)}_L= \tanh (\varepsilon/2T_{1,2})$ 
  characterized by the different temperatures $T_{1,2}$ in the electrodes S$_{1,2}$.   
 
 {\color{black}
 Proximity of the outer ferromagnetic layers shown in Fig.(\ref{Fig:SFSmodel})
 induces Zeeman splitting of electronic states which acts as an effective exchange field 
 in the superconducting electrodes. We assume that superconducting layers are thin enough  to neglect the 
 spatial variations of the spectral GFs (retarded and advances)
 so that up to leading order they retain their bulk values in the presence of a homogeneous exchange field
 }
 \begin{equation} \label{Eq:GR}
 g^R = \tau_{3} \left[ g_{03} + g_{33} (\bm \sigma \bm h) \right] +
       \tau_{1}\left[ g_{01} + g_{31} (\bm \sigma \bm h) \right]
  \end{equation}
 and $ g^A=-\tau_3 g^{R\dag}\tau_3$.
 The terms diagonal in Nambu space ($\tau_3$) correspond to the normal correlations 
 which determine the total density of states (DOS) and the 
 DOS difference between the spin-up and spin-down subbands is given by the components
 $N_+ = {\rm Re} g_{03}$ and $N_- = {\rm Re} g_{33}$ respectively. 
 The  off-diagonal components ($\tau_1$) describe spin-singlet $g_{01}$ and spin-triplet   
 $g_{31}$ superconducting correlations which appears as the result of the exchange splitting  \cite{Bergeret2001}. 
    
  The tunnelling heat current $\dot Q$ across the Josephson junction (JJ) is given by the general expression
  \begin{equation} \label{Eq:Q}
  R_N \dot Q = \frac{1}{16 e^2} \int_{\infty}^{\infty} d\varepsilon \varepsilon \;{\rm Tr}( \check{I}^K_{12})  \\
  \end{equation}
   where $R_N$  is the normal-state resistance of the tunnelling barrier and $e$ is the electron charge. 
   At first we calculate the heat current using (\ref{Eq:Q}). Assuming that the temperature difference is small 
 we expand the distribution functions in S$_{1,2}$ electrodes 
 $f_{L}^{(1,2)}  = f_{0} + (T_{1,2}-T) \frac{\partial f_0}{\partial T}$
 where $T= (T_1+T_2)/2$ and introduce the total heat conductance as $ \kappa = \dot Q / (T_1 - T_2)$. 
 In accordance with the Eq.(\ref{Eq:Q}) it can be written as the 
 superposition of three terms  
 \begin{equation}\label{Eq:kappaGen}
 \kappa/\kappa_N = \kappa_{qp} - \kappa_{c}\cos\theta - \kappa_s  \sin\theta,
 \end{equation}
 where $\kappa_N =\pi^2 T/(3e^2R_N) $ is the normal state thermal
  conductance of the junction. We will refer to the different contributions in (\ref{Eq:kappaGen})  as 
the quasiparticle $\kappa_{qp}$, the usual $\kappa_c$ and phase-shifting $\kappa_{s}$ interference terms. 
Assuming S$_{1,2}$ superconductors to be identical we get expressions for components in
 Eq.(\ref{Eq:kappaGen}):
 \begin{align} \label{Eq:kappaQP}
 &\kappa_{qp} = \int_{-\infty}^{\infty} d\varepsilon F 
 \{ N_+^2 + [r ({\bm h}_{1\perp} {\bm h}_{2\perp}) +  h_{1\parallel}h_{2\parallel} ] N_-^2 \} ,
 \\  \label{Eq:kappaC} 
 & \kappa_{c} =   \int_{-\infty}^{\infty} d\varepsilon  F \times  \\  \nonumber
 &  \{ r ({\rm Im} g_{01})^2 + \left[ r h_{1\parallel}h_{2\parallel} + ({\bm h}_{1\perp} {\bm h}_{2\perp})  \right]
 ({\rm Im}g_{31})^2 \} ,
 \\  \label{Eq:kappaS}
   &\kappa_{s} = \chi P \int_{-\infty}^{\infty} d\varepsilon  F ({\rm Im} g_{31})^2 ,
 \end{align}
  where  $\chi = {\bm m}\cdot (\bm h_1 \times \bm h_2)$ is spin chirality,  $r=\sqrt{1-P^2}$,  
  $h_\parallel = ({\bm m}{\bm h})$ and
 ${\bm h}_{\perp} = {\bm h} - h_\parallel {\bm m} $
 are the exchange field components parallel and perpendicular to the FI barrier polarization,
   $F(\varepsilon) = 
(6\varepsilon^2/\pi^2 T^2) \partial f_0/ \partial \varepsilon $.
  
 %%%%%%%%%%%%%%%%%%%%%%%%%%%%%%%%%%%%%%%%%%%%%%%%
 \begin{figure}[htb!]
 \centerline{$
 \begin{array}{c}
 \includegraphics[width=1.0\linewidth]{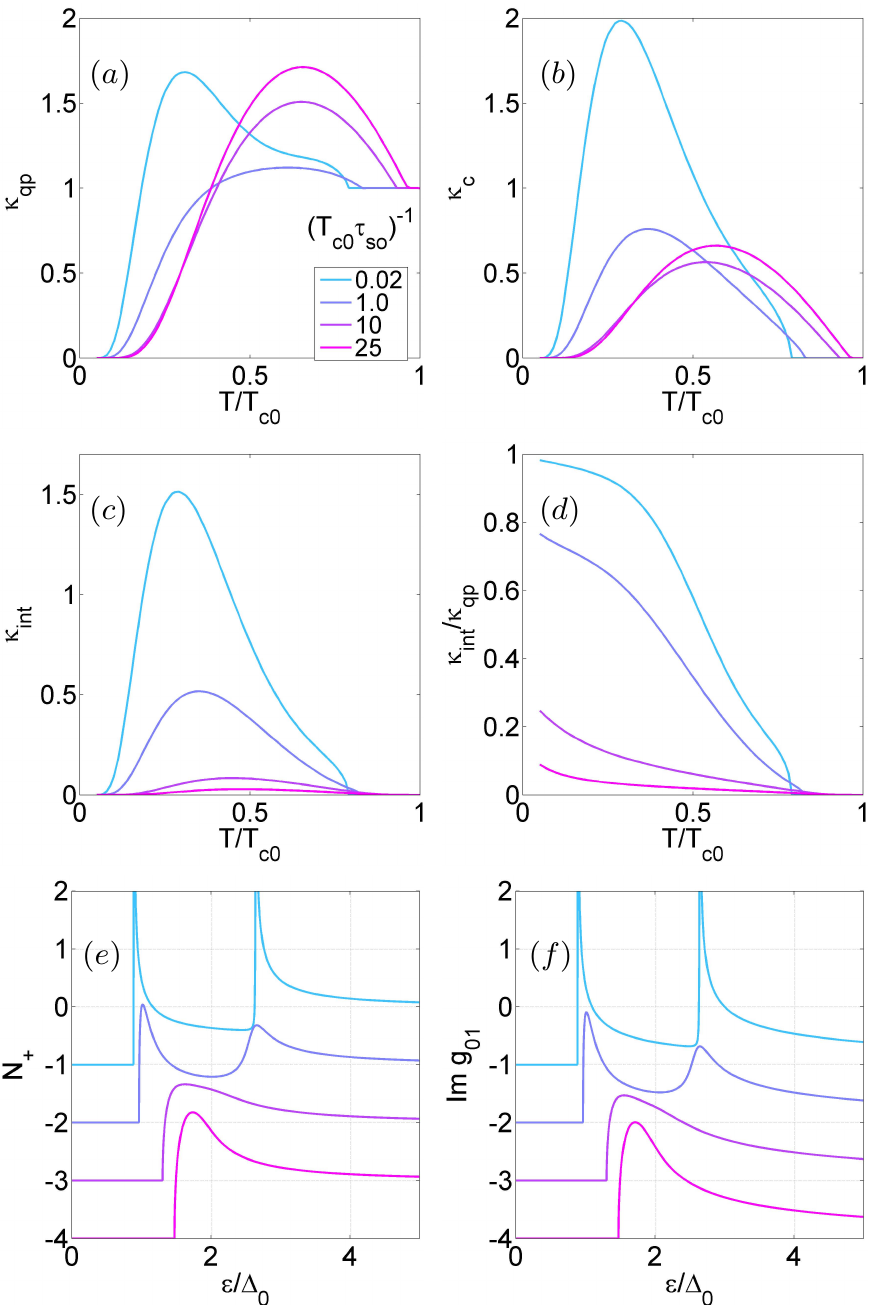}
  \end{array}$}
 \caption{\label{Fig:3} (Color online) 
 Temperature dependencies of heat conductance contributions (a) $\kappa_{qp}$,  (c) $\kappa_{int}$
  and their ratio (d) $\kappa_{int}/\kappa_{qp}$ 
 for the SFS structure in Fig.(\ref{Fig:SFSmodel}) 
 with ${\bm h}_{1,2} \perp {\bm m} $ and the 100\% spin filtering $P=1$. 
 (b) $\kappa_{c}$ for the same structure but with $P=0.8$ and $\theta_h = \pi/4$. 
 (e) Total DOS $N_+(\varepsilon)$ and (f) spin-singlet  anomalous function ${\rm Im} g_{01} (\varepsilon)$. 
 The exchange field is $h=0.5\Delta_0$ and the values
   of spin relaxation rate $(T_{c0}\tau_{so})^{-1}$ shown in (a) are the same for all panels.
     }
 \end{figure}   
 %%%%%%%%%%%%%%%%%%%%%%%%%%%%%%%%%%%%%%%%%%%%%%%%%%%  
  
   The quasiparticle and the usual phase-sensitive contributions
   (\ref{Eq:kappaQP},\ref{Eq:kappaC})  have been analysed for the coplanar magnetic configuration \cite{Bergeret2013}.  
  The term  $\kappa_s$ (\ref{Eq:kappaS}) is non-zero only in the non-coplanar
  case $\chi\neq 0$ and it produces the phase shift of HCPR  in Eq.(\ref{Eq:HCPRgen}) 
   given by $\theta_0 =  \arctan (\kappa_s / \kappa_c)$. 
    Comparing  different parts of the conductance (\ref{Eq:kappaQP},\ref{Eq:kappaC},\ref{Eq:kappaS}) one can see 
    that $\kappa_s$ is qualitatively different from the others since it stems exclusively from the triplet part
     of the condensate associated with the GF component $g_{31}$. In the general case of a non-ideal spin-filter $P\neq 1$ 
     the usual interference  part $\kappa_{c}$ has contributions from both spin-singlet and spin-triplet Cooper pairs.  
      Thus one can conclude that the non-trivial phase shift $\theta_0\neq 0$ of the HCPR is a direct experimentally measurable 
    evidence of the transport of spin-triplet Cooper pairs across the tunnel junction.   
    
     Physically the phase-shifting term  $\kappa_s$ appears as a result of the additional phase picked up by the spin-triplet 
     Cooper pairs when tunnelling  between two superconductors with non-collinear exchange fields through the spin-polarising barrier. 
     To understand this phenomenon 
     on a qualitative level let us consider the  magnetic configuration
     ${\bm h_1} = h_1 {\bm z} $, ${\bm h_2} = h_2 {\bm x}$ and ${\bm m} =  {\bm y}$. 
     The spin-triplet condensates in S$_1$ and S$_2$ are described by the  wave functions      
      $ \Psi_{t1} \sim |\uparrow,\downarrow\rangle_z  + |\downarrow,\uparrow\rangle_z$
     and  $ \Psi_{t2} \sim e^{i\varphi}(|\uparrow,\downarrow\rangle_x  + |\downarrow,\uparrow\rangle_x)$ respectively, 
     where  the spin quantization axes are set by the directions of exchange fields ${\bm h_{1,2}}$.
    Assuming  the spin filter to be ideal $P=1$ we find for the tunnelling amplitude 
    $\langle \Psi_{2t} |\hat P_y |\Psi_{1t} \rangle \sim i e^{i\varphi}$, where $\hat P_y$ is the projection operator 
    acting on each of the single-electron states $\hat P_y |\uparrow\rangle_{z} = \frac{1}{2}|\uparrow\rangle_{y}$ and 
    $\hat P_y |\downarrow\rangle_{z} = \frac{-i}{2} |\uparrow\rangle_{y}$.
    Thus one can see that the spin-filtering provides an additional $\pi/2$ phase in the tunnelling amplitude of spin-triplet
     Cooper pairs, which is the origin of the anomalous Josephson effect \cite{Silaev2017} and the phase-shifted HCPR studied here.  
    
      On a quantitative level let us analyse the particular configuration  ${\bm h}_{1,2}\perp {\bm m}$ when 
  the expressions (\ref{Eq:kappaC},\ref{Eq:kappaS}) yield the 
  interference contributions which are proportional to each other $\kappa_{int} = 
  \int_{-\infty}^{\infty} d\varepsilon F(\varepsilon)   ({\rm Im} g_{31})^2  $ so that 
  $\kappa_{c} = -\cos\theta_{h} \kappa_{int}$ and    $\kappa_{s} = \sin\theta_{h} \kappa_{int}$,
   where $\theta_{h}$ is the angle between ${\bm h_{2\perp}}$ and ${\bm h_{1\perp}}$ shown schematically in Fig.(\ref{Fig:SFSmodel}).
  Hence for the ideal spin-filter $P=1$ the phase shift of HCPR is determined by the geometry of magnetic configuration 
  $\theta_0 = \theta_h$
  although the overall amplitude of $\kappa_{int}$ is strongly suppressed by the spin relaxation. 
  To demonstrate this  we find the order parameter and spectral functions (\ref{Eq:GR})
   self-consistently taking onto account the presence of exchange field and SO scattering rate which can vary in wide limits corresponding to\cite{Poli2008} 
   $(\Delta_0 \tau_{so})^{-1} \approx 0.2$ in Al and to \cite{Wakamura2014} $(\Delta_0 \tau_{so})^{-1} \approx 500$ in Nb,
    where $\Delta_0$ is the 
   bulk superconducting gap at  $h=0$,  $\tau_{so}=\infty$ and $T\to 0$.

     %%%%%%%%%%%%%%%%%%%%%%%%%%%%%%%%%%%%%%%%%%%%%%%%
 \begin{figure}[htb!]
 \centerline{$
 \begin{array}{c}
 \includegraphics[width=1.0\linewidth]{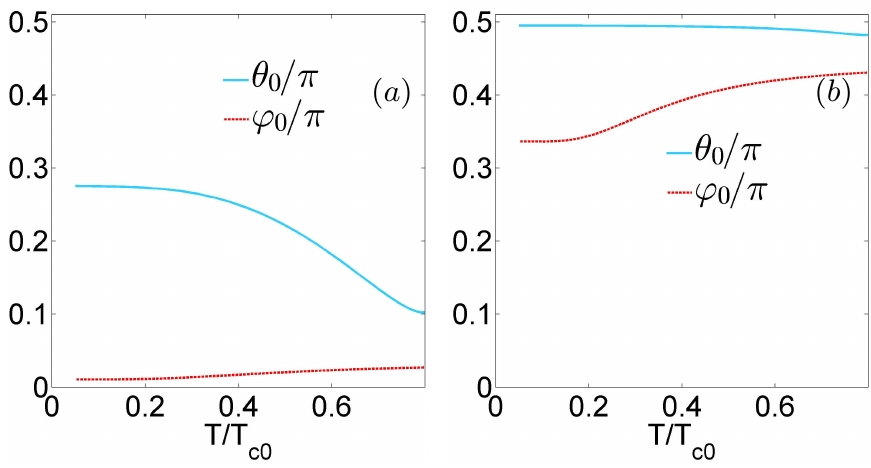}
  \end{array}$}
 \caption{\label{Fig:4} (Color online) 
 The phase shifts of CPR $\varphi_0 = \varphi_0(T) $ and HCPR  $\theta_0 = \theta_0(T) $
 for (a) $P=0.8$ and (b) $P=0.9999$.  Exchange splitting $h=0.5\Delta_0$ and SO relaxation $(T_{c0}\tau_{so})^{-1} =1$.
  The magnetic configuration is that  ${\bm h}_{1,2}\perp {\bm m}$ and $\theta_h = \pi/2$.    }
 \end{figure}   
 %%%%%%%%%%%%%%%%%%%%%%%%%%%%%%%%%%%%%%%%%%%%%%%%%%%       
 
   Let us consider the calculated dependencies of quasiparticle $\kappa_{qp} (T) $ and interference 
   $\kappa_{int} (T)$ parts  shown in Figs. (\ref{Fig:3})a and (\ref{Fig:3})c respectively for the 
   fixed exchange field $h=0.5\Delta_0$.
      First there is a non-monotonic dependence $\kappa_{qp} = \kappa_{qp}(T)$ which increases above the normal 
    state value  as the temperature goes down below $T_c$.
     This behaviour is explained by the  DOS enhancement near the gap edge.
    At lower temperatures $T\ll T_c$ the quasipartiles are frozen out which results in the exponential drop of the 
     heat conductance. Of interest is the evolution of the peak amplitude in the $\kappa_{qp}(T)$ dependence with increasing 
     spin relaxation rate. Initially  with increasing $\tau_{so}^{-1}$ from zero to small values the peak of $\kappa_{qp}(T)$ is suppressed
     while at the larger values of $\tau_{so}^{-1}$ it is restored.  This tendency
     reveals the 
     the evolution of DOS $N_+ (\varepsilon)$ and the anomalous function 
     ${\rm Im}g_{01}(\varepsilon) $ with increasing $\tau_{so}^{-1}$ shown in Fig.(\ref{Fig:4}). 
     For  $\tau_{so}^{-1} \ll T_{c0}$ the singularities of spectral functions are smeared. 
     However at larger values of $\tau_{so}^{-1} > T_{c0}$ 
   both $N_+$ and ${\rm Im}g_{01}$ again develop the peaks although without the spin-splitting features. 
   At the same time the spin-triplet components $g_{33}$ and $g_{13}$ are strongly suppressed by the SO scattering.   
     
  The temperature dependence of interference thermal conductance $\kappa_{int}(T)$ is also non-monotonic
  due to the similar mechanism as discussed above. As shown in Fig.(\ref{Fig:3})c 
  the maximum of $\kappa_{int}(T)$ is strongly suppressed by the SO scattering
      which tends to remove the spin-dependent components $g_{31}$ of the GFs. 
   However for the    fully-polarizing ideal spin filter $P=1$ according to the equitation (\ref{Eq:kappaC})
   the suppression  of $\kappa_{int}$ does not affect the phase shift of HCPR which is fixed by the angle between 
   exchange fields ${\bm h_{1,2}}$ as discussed above. 
    
    The situation is different for $P<1$ when $\kappa_c$ and $\kappa_s$ are no longer proportional to each other. 
    The qualitative difference between the usual $\kappa_c$ and phase-shifting $\kappa_s$
    contributions in this case is determined by the transport of spin-singlet Cooper pairs which is only possible
    if both the spin projections can pass the spin filter. 
    This contribution is described by the first term in the r.h.s of Eq.(\ref{Eq:kappaC}) 
     which yields a non-zero contribution since $r\neq 0$.     
    In this case the phase shift of HCPR is suppressed by the SO scattering
    which leads to the decrease of spin-triplet correlations so that $\kappa_s\to 0$.
   At the same time the contribution of spin-singlet Cooper pairs survives keeping 
    $\kappa_c\neq 0$ in the limit $\tau_{so} \to 0$ as shown in the Fig.(\ref{Fig:3})b.       
    
  It is instructive to compare the phase-shifted HCPR (\ref{Eq:HCPRgen}) and 
  CPR   (\ref{Eq:CPRgen}) calculated for the spin-valve shown in Fig.(\ref{Fig:SFSmodel}) 
  using the general matrix current (\ref{Eq:KupLuk}).   
  The usual $I_0 = I_c \cos\varphi_0$ and anomalous $I_{an} = I_c \sin\varphi_0$ 
 Josephson currents through tunnel barrier are given by
   \begin{align} \label{Eq:I0}
   & \frac{R_N I_0}{\pi e T} =  \sum_{\omega_n} \left[ r (g_{01}^2 + 
    h_{1\parallel}h_{2\parallel} g_{31}^2) + ({\bm h_{1\perp}}{\bm h_{2\perp}}) g_{31}^2  \right]
    \\ \label{Eq:Ian}
   & \frac{R_N I_{an}}{\pi e T} = \chi P\sum_{\omega_n}  g_{31}^2,
   \end{align}
  where $\chi = {\bm m} ({\bm h_1\times \bm h_2})$ is the spin chirality,  
 $g_{01}$ and $g_{31}$ are the spin-singlet and spin-triplet components of the Matsubara GF written in the form (\ref{Eq:GR}) analytically continued   to the imaginary frequencies $\varepsilon \to i\omega_n$ with $\omega_n = (2n+1)\pi T$. 

 Expressions for the Josephson current (\ref{Eq:I0},\ref{Eq:Ian}) are dual to that of the
 interference heat conductance (\ref{Eq:kappaC},\ref{Eq:kappaS}). Similar to  the phase-shifting term $\kappa_s$
 the anomalous current $I_{an}$ is mediated by spin-triplet component $g_{31}$. 
 Therefore $\varphi_0$ Josephson effect is the directly observable signature of the spin-triplet superconducting current
  across the junction. 
For the ideal spin filter  $P=1$  Eqs.(\ref{Eq:I0},\ref{Eq:Ian}) yield a temperature-independent
 phase shift of CPR $\varphi_0=\theta_0 = \theta_h$, although in the general case $\theta_0$ and $\varphi_0$
 can be quite different, as shown in Fig.(\ref{Fig:4})a,b.  
     
%   {\bf Experimental observations} 
   The predicted effect of phase-shifted HCPR can be experimentally observed 
    using the Josephson heat interferometer \cite{Giazotto2012,Martinez-Perez2014} consisting of the 
   temperature-biased superconducting quantum interference device (SQUID ) with 
    the usual  JJ in the one part and $\theta_0$-JJ in the other part. 
     In this case  following the derivation in Ref.(\onlinecite{Martinez-Perez2014})
   one can show that the interference pattern of the heat current across the SQUID 
   $\dot Q_{int}$
   contains a spontaneous shift as a function of the external  magnetic flux $\Phi$ so that
   $\dot Q_{int} = \dot Q_{int} ( \Phi - \Phi_e) $. 
  For the ideal spin filter when $\varphi_0=\theta_0 = \theta_h$ we get   
    $\Phi_e = \theta_h \Phi_0/2\pi$
  
  The $\theta_0$-shifted HCPR (\ref{Eq:CPRgen}) provides an 
  interesting possibility to couple the heat transport with magnetization dynamics. 
 Oscillations of  moments ${\bm h_{1,2}}$ and ${\bm m}$ driven by the Larmour precession around the effective field\cite{Konschelle2009}
  in the generic thermomagnetic circuit 
   Fig.(\ref{Fig:SFSmodel}) produce the time-dependent spin chirality $\chi = \chi (t)$ and hence 
 generate the non-stationary phase shifts $\varphi_0 = \varphi_0 (t)$ and $\theta_0 = \theta_0 (t)$.
 Thus   according to Eqs.(\ref{Eq:HCPRgen},\ref{Eq:CPRgen}) one can generate alternating heat and charge currents   
    at the Larmour frequency  which can be controlled by an external magnetic an the anisotropy field. 
   The other possible application is based on the effective  conversion of spin currents inside the ferromagnet or ferromagnetic insulator 
   into the electronic heat and charge currents across the attached Josephson junctions.
   Based on the discussed effect it is in principle possible to implement the  superconducting JJ detector of magnetic precession 
  associated with magnons in FI layer  \cite{Satoh2010,Uchida2010, Bauer2012}  or the skyrmion motion inside ferromagnets 
  which can be used for the racetrack magnetic memory applications\cite{Fert2013}. 
   
 To summarize, 
 we have found the thermal counterpart of the anomalous Josephson effect. 
 Under the conditions of 
 broken time-reversal and chiral symmetries the interference heat current acquires an arbitrary phase shift $\theta_0$
 which substantially generalizes the previously found forms of HCPR. 
 For the generic example of the non-coplanar Josephson spin valve (Fig.\ref{Fig:SFSmodel})
 $\theta_0$ is determined by the non-zero spin chirality $\chi = {\bm m}\cdot (\bm h_1 \times \bm h_2) \neq 0$. 
 The phase shift is demonstrated to be the direct and experimentally measurable evidence of the heat transport
 with participation of spin-triplet Cooper pairs.  
   In view of possible applications the proposed effect allows to change the heat conductance of the system 
 in a continuous way by rotating magnetic vectors. 
 For this purpose it is preferable to use magnetic elements  with different 
 coercivity fields \cite{Li2013} or anisotropies.    
   
 %{\bf Acknowledgements}
 We thank T. T Heikkil\"a , F.S. Bergeret, A. Mel'nikov, I. Bobkova and A. Bobkov
 for stimulating discussions.
  The work was supported by the Academy of Finland.

\end{document}